\documentclass{llncs}

\usepackage{amsfonts,amsmath,amssymb}
\usepackage{xspace}
\usepackage{hyperref,paralist}
\usepackage{multirow}
\usepackage{graphicx}

\usepackage{caption}
\usepackage{subcaption}

\usepackage{breakurl} 

\usepackage{listings}
\usepackage{pstricks,pst-node,pst-plot,pstricks-add}

\newcommand{\submission}[1]{{}} 
\newcommand{\final}[1]{{#1}} 
\newcommand{\rronly}[1]{{#1}} 
\newcommand{\pponly}[1]{{}} 

\renewcommand{\circledR}{{}}

\newcommand{\trans}{T}

\newcommand{\slice}{\mathit{slice}}

\newcommand{\cbmc}{\textsc{Cbmc}\xspace}
\newcommand{\btc}{\textsc{BTC}\xspace}
\newcommand{\btcet}{\textsc{Em\-bed\-ded\-Tes\-ter$^\circledR$}\xspace}
\newcommand{\btcev}{\textsc{Em\-bed\-ded\-Va\-li\-da\-tor$^\circledR$}\xspace}
\newcommand{\mlsl}{\textsc{\rronly{MATLAB$^\circledR$ }Simulink$^\circledR$}\xspace}
\newcommand{\stateflow}{\textsc{Stateflow$^\circledR$}\xspace}
\newcommand{\simulink}{\textsc{Simulink$^\circledR$}\xspace}
\newcommand{\dstl}{\rronly{dSPACE$^\circledR$ }\textsc{TargetLink$^\circledR$}\xspace}
\newcommand{\targetlink}{\textsc{TargetLink$^\circledR$}\xspace}
\newcommand{\btces}{BTC-ES\xspace}
\newcommand{\btcetf}{\rronly{\btcet}\pponly{test case generator}\xspace}
\newcommand{\btcevf}{\rronly{\btcev}\pponly{requirements validator}\xspace}

\newcommand{\fuelsys}{\textsc{FuelSys}\xspace}

\newcommand{\minisat}{\textsc{MiniSAT2}\xspace}

\renewcommand{\paragraph}[1]{\noindent\textbf{#1}.~}

\begin{document}

\title{Incremental Bounded Model Checking\\ for Embedded Software
  \rronly{(extended version)}%
\thanks{The research leading to these results has received funding
    from the ARTEMIS Joint Undertaking under grant
    agreement number 295311 \href{http://vetess.eu/}{``VeTeSS''} and ERC project 280053.}}
\author{Peter Schrammel\inst{1} \and Daniel Kroening\inst{1} \and Martin Brain\inst{1} \and Ruben Martins\inst{1} \and\\ Tino Teige\inst{2} \and Tom Bienm\"uller\inst{2}}
\institute{University of Oxford \and BTC Embedded Systems AG}

\maketitle
\begin{abstract}
Program analysis is on the brink of mainstream in embedded systems
development.  Formal verification of behavioural requirements, finding
runtime errors and automated test case generation are some of the most
common applications of automated verification tools based on Bounded Model
Checking.
Existing industrial tools for embedded software use an off-the-shelf
Bounded Model Checker and apply it iteratively to verify the program
with an increasing number of unwindings.  This approach unnecessarily
wastes time repeating work that has already been done and fails to
exploit the power of incremental SAT solving.
This paper reports on the extension of the software model checker
\cbmc to support \emph{incremental Bounded Model Checking} and its
successful integration with \submission{an}\final{the} industrial
embedded software verification tool\final{ \btc \btcet}.
We present an extensive evaluation over large industrial embedded programs,
which shows that incremental Bounded Model Checking cuts runtimes by
\emph{one order of magnitude} in comparison to the standard non-incremental
approach, enabling the application of formal verification to large and
complex embedded software.
\end{abstract}

\section{Introduction}\label{sec:intro}

Recent trend estimation~\cite{GKF+12} in automotive embedded systems
revealed ever growing complexity of computer systems, providing increased
safety, efficiency and entertainment satisfaction.
Hence, automated design tools are vital for managing this complexity and
support the verification processes in order to satisfy the high safety
requirements stipulated by safety standards and regulations.
Similar to the developments in hardware verification in the 1990s,
verification tools for embedded software are becoming indispensable
in industrial practice for hunting runtime bugs, checking 
functional properties and test suite generation~\cite{FWA09}.
For example, the automotive safety standard ISO 26262~\cite{ISO26262}
requires the test suite to satisfy modified condition/decision
coverage~\cite{HVCR01} -- a goal that is laborious to achieve without
support by a model checker that identifies unreachable test goals and
suggests test vectors for difficult-to-reach test goals.

In this paper, we focus on the application of Bounded Model Checking
(BMC) to this problem.  The technique is highly accurate (no false
alarms) and is furthermore able to generate counterexamples that aid
debugging and serve as test vectors.
The spiralling power of SAT solvers has made this technique scale to
reasonably large programs and has enabled industrial application.

In BMC, the property of interest is checked for traces that execute loops up
to a given number of times $k$.  Since the value of $k$ that is required to
find a bug is not known a-priori, one has to try increasingly larger values
of $k$ until a bug is found.  The analysis is aborted when memory and
runtime limits are exceeded.%
\footnote{One can stop unwinding when the
\emph{completeness
threshold}~\rronly{\cite{KS03,KOS+11}}\pponly{\cite{KS03}} of the system is
reached, but this threshold is often impractically large.}

Most existing industrial verification tools use an off-the-shelf
Bounded Model Checker and, without additional information about the
program to be checked, apply it in an iterative fashion:

{
\footnotesize
\lstset{language=sh,emph={od},emphstyle={\bfseries}}
\begin{lstlisting}
k=0
while true do
  if BMC(program,k) fails then
    return counterexample
  fi
  k++
od
\end{lstlisting}
\lstset{language=C}
}

This basic procedure offers scope for improvement. In particular, note that
the Bounded Model Checker has to redo the work of generating and solving the
SAT formula for time frames $0$ to $k$ when called to check time frame
$k+1$.  It is desirable to perform the verification \emph{incrementally} for
iteration $k+1$ by building upon the work done for iteration~$k$.

Incremental BMC has been applied successfully to the verification of
hardware designs, and has been reported to yield substantial
speedups~\cite{Str01,ES03b}.  Fortunately, the typical control-loop
structure of embedded software resembles the monolithic transition relation
of hardware designs, and thus strongly suggests incremental verification of
successive loop unwindings.  However -- to our knowledge -- none of the
software model checkers for C programs that have competed in the TACAS 2014
Software Verification Competition implement such a technique that ultimately
exploits the full power of incremental SAT solving~\cite{WKS01,ES03a}.

\paragraph{Contributions}
The primary contribution of this paper is \emph{experimental}. We quantify
the benefit of incremental BMC in the context of the verification
of embedded software. To this end,
\begin{compactenum}[(1)]
\item we survey the techniques that are
  state of the art in embedded software verification, briefly
  summarise the underlying theory, and highlight the challenges 
  faced when applying them to industrial code;
\item we present the first industrial-strength implementation of
  incremental BMC in a software model checker for ANSI-C programs
  combining symbolic execution, slicing and incremental SAT solving.
  Besides loop unwinding, we also elucidate other applications of
  incremental SAT solving in a state-of-the-art Bounded Model Checker;
\item we report on the successful integration of our incremental
  bounded model checker in the industrial embedded software
  verification tools \btc \btcet and \btcev where it is used by
  several hundred industrial users since version 3.4 and 4.3,
  respectively; and
\item we give a comprehensive experimental evaluation over a large set of
  industrial embedded benchmarks that quantify the
  performance gain due to incremental BMC: our new method
  outperforms the winner of the TACAS 2014 Software
  Verification Competition~\cite{KT14} by one order of magnitude.
\end{compactenum}

\section{Verification of Model-based Embedded Software}\label{sec:appl}

Model-based development is well-est\-abli\-shed in embedded software
development, and particularly popular in the automotive industry.
Tools such as
\simulink\rronly{\footnote{\url{http://www.mathworks.co.uk/products/simulink/}}}
are in widespread use for modelling, code generation and testing.%
\footnote{The topic of model-based testing methods is discussed
in detail in a range of surveys~\cite{CRT10,DT10,PSM12}.}

In this paper, we focus on the verification of C code generated from these
models.  To this end, we illustrate the characteristics of this verification
problem with the help of a well-known case
study\rronly{ (Sec.~\ref{sec:casestudy})} and explain the workflow and principal
techniques that a state-of-the-art embedded software verification tool uses.

\subsection{Case Study: Fault-Tolerant Fuel Control System}\label{sec:casestudy}

\begin{figure*}[t]
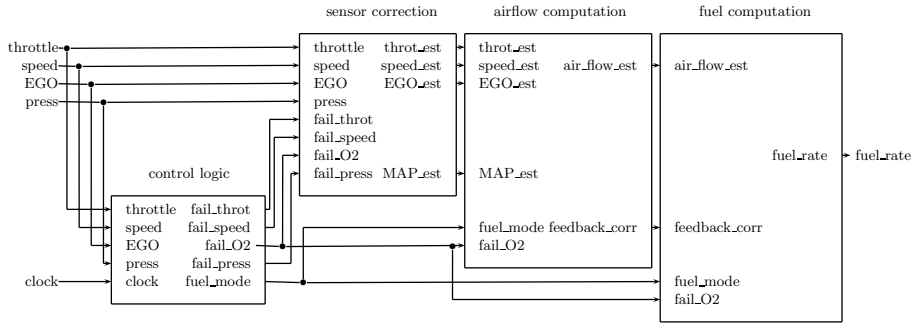

\scalebox{0.62}{
\begin{tabular}{rcccc@{\;}lcrcccc@{\;\;}lr@{\;\;}lr@{\;}c@{\;}lrr@{\;\;}cc}
  & & & & & & & & & & & & \multicolumn{2}{c}{sensor correction} & \multicolumn{2}{c}{airflow computation} & & \multicolumn{3}{c}{fuel computation} \\
  & & & & & & & & & & & & \pnode{sclu} & \pnode{scru} & \pnode{aclu} & \pnode{acru} & & \pnode{fclu} & & \pnode{fcru} \\
\Rnode{t0}{throttle} & \rnode{t12}{$\bullet$} & & & & & & & & & & & \Rnode{t1}{} \; throttle & throt\_est \; \Rnode{te0}{} & \Rnode{te1}{} \; throt\_est & & & & & & & \\
\Rnode{s0}{speed} & & \rnode{s12}{$\bullet$} & & &  & & & & & & & \Rnode{s1}{} \; speed & speed\_est \; \Rnode{se1}{} & \Rnode{se2}{} \; speed\_est & air\_flow\_est \; \Rnode{ae1}{} & & \Rnode{ae2}{} \; air\_flow\_est & & & & \\
\Rnode{e0}{EGO}     & & & \rnode{e12}{$\bullet$} & &  & & & & & & & \Rnode{e1}{} \; EGO & EGO\_est \; \Rnode{ee1}{} & \Rnode{ee2}{} \; EGO\_est & & & & & & & \\
\Rnode{p0}{press}   & & & & \rnode{p12}{$\bullet$} & & & & & & & & \Rnode{p1}{} \; press && & & & & & & & \\
& & & & & & & & \Rnode{ft11}{} & & & & \Rnode{ft1}{} \; fail\_throt & & & & & & & & & \\
& & & & & & & & & \Rnode{fs11}{} & & & \Rnode{fs1}{} \; fail\_speed && & & & & & & & \\
& & & & & & & & & & \Rnode{fo11}{} & & \Rnode{fo1}{} \; fail\_O2 && & & & & & fuel\_rate \; \Rnode{fr1}{} & \Rnode{fr2}{} fuel\_rate \\
& & & & & \multicolumn{3}{c}{control logic} & & & & \Rnode{fp11}{} & \Rnode{fp1}{} \; fail\_press & MAP\_est \; \Rnode{me1}{} & \Rnode{me2}{} \; MAP\_est & & & & & & & \\
& & & & & \pnode{cllu} && \pnode{clru} & & & & & \pnode{scll} & \pnode{scrl}  \\
& \Rnode{t22}{} & &  & & \Rnode{t2}{} \; throttle & & fail\_throt \; \Rnode{ft2}{} & \Rnode{ft22}{} & &  & & & & & & & & & \\
& & \Rnode{s22}{} & & & \Rnode{s2}{} \; speed & & fail\_speed \; \Rnode{fs2}{} & & \Rnode{fs22}{} & & & \hspace*{0.25em}\Rnode{fm11}{} & & \Rnode{fm1}{} \; fuel\_mode & feedback\_corr \; \Rnode{fc1}{} & & \Rnode{fc2}{} \; feedback\_corr & & & & \\
& & & \Rnode{e22}{} & & \Rnode{e2}{} \; EGO && fail\_O2 \Rnode{fo4}{} \; & & & \rnode{fo12}{$\bullet$} & & & \rnode{fo23}{$\bullet$} & \Rnode{fo3}{} \; fail\_O2 & & & & & & & \\
& & & & \Rnode{p22}{}  & \Rnode{p2}{} \; press & & fail\_press \; \Rnode{fp2}{} & & & & \Rnode{fp22}{} & & & \pnode{acll} & \pnode{acrl} & \\
\Rnode{c0}{clock}   & & & & & \Rnode{c1}{} \; clock & & fuel\_mode \; \Rnode{fm4}{} & & & & & \rnode{fm12}{$\bullet$} &  & & & & \Rnode{fm6}{} \; fuel\_mode \\
& & & & & \pnode{clll} && \pnode{clrl} & & & & & & \Rnode{fo33}{}\hspace*{0.25em} & & & & \Rnode{fo5}{} \; fail\_O2 \\
 & & & & & & & & & & & & & & & & & \pnode{fcll} && \pnode{fcrl} \\
\end{tabular}
\ncline{cllu}{clru} \ncline{clru}{clrl} \ncline{clrl}{clll} \ncline{clll}{cllu}
\ncline{sclu}{scru} \ncline{scru}{scrl} \ncline{scrl}{scll} \ncline{scll}{sclu}
\ncline{aclu}{acru} \ncline{acru}{acrl} \ncline{acrl}{acll} \ncline{acll}{aclu}
\ncline{fclu}{fcru} \ncline{fcru}{fcrl} \ncline{fcrl}{fcll} \ncline{fcll}{fclu}
\ncline{t0}{t12}\ncline{t12}{t22}\ncline[arrows=->]{t12}{t1}\ncline[arrows=->]{t22}{t2}
\ncline{p0}{p12}\ncline{p12}{p22}\ncline[arrows=->]{p12}{p1}\ncline[arrows=->]{p22}{p2}
\ncline{s0}{s12}\ncline{s12}{s22}\ncline[arrows=->]{s12}{s1}\ncline[arrows=->]{s22}{s2}
\ncline{e0}{e12}\ncline{e12}{e22}\ncline[arrows=->]{e12}{e1}\ncline[arrows=->]{e22}{e2}
\ncline[arrows=->]{te0}{te1}
\ncline[arrows=->]{se1}{se2}
\ncline[arrows=->]{ae1}{ae2}
\ncline[arrows=->]{ee1}{ee2}
\ncline[arrows=->]{me1}{me2}
\ncline[arrows=->]{fo1}{fo2}
\ncline[arrows=->]{fr1}{fr2}
\ncline[arrows=->]{c0}{c1}
\ncline[arrows=->]{fo5}{fo6}
\ncline[arrows=->]{fc1}{fc2}
\ncline{ft2}{ft22}\ncline{ft22}{ft11}\ncline[arrows=->]{ft11}{ft1}
\ncline{fs2}{fs22}\ncline{fs22}{fs11}\ncline[arrows=->]{fs11}{fs1}
\ncline{fp2}{fp22}\ncline{fp22}{fp11}\ncline[arrows=->]{fp11}{fp1}
\ncline{fo4}{fo12}\ncline{fo12}{fo11}\ncline{fo23}{fo33}\ncline[arrows=->]{fo11}{fo1}\ncline[arrows=->]{fo12}{fo3}\ncline[arrows=->]{fo33}{fo5}
\ncline{fm4}{fm12}\ncline{fm12}{fm11}\ncline[arrows=->]{fm11}{fm1}\ncline[arrows=->]{fm12}{fm6}
}
\caption{\label{fig:fuelsys}
The \simulink Diagram for the Fault-Tolerant Fuel Control System}
\end{figure*}

The Fault-Tolerant Fuel Control
System\footnote{\url{http://www.mathworks.co.uk/help/simulink/examples/modeling-a-fault-tolerant-fuel-control-system.html}}
(\fuelsys) for a gasoline engine\rronly{, originally introduced as a
demonstration example for \mlsl/\stateflow and then ad\-apted for
\dstl,} is representative of a variety of automotive applications
as it combines discrete control logic \rronly{via \stateflow }with
continuous signal flow\rronly{ expressed by \simulink or \targetlink}
and thus establishes a hybrid discrete-continuous system.  More
precisely, the control logic of \fuelsys is implemented by six
automata with two to five states each, while the signal flow is
further subdivided into three subsystems with a rich variety of
\simulink\final{/\targetlink} blocks involving arithmetic, lookup tables,
integrators, filters and interpolation (Fig.~\ref{fig:fuelsys}).
The system is designed to keep the air-fuel ratio nearly constant
depending on the inputs given by a throttle sensor, a speed sensor, an
oxygen sensor (EGO) and a pressure sensor (MAP).  Moreover it is
tolerant to individual sensor faults and is designed to be highly
robust, i.e.~after detection of a sensor fault the system is dynamically
reconfigured.

\paragraph{Properties of interest}
The key functional property for \fuelsys is how the air-fuel ratio evolves
for each of the four sensor-failure scenarios.  Simulation-based approaches
show that \fuelsys is indeed fault-tolerant in each case of a single
failure: the air-fuel ratio can be regulated after a few seconds to about
$80\,\%$ of the target ratio.
In addition to \emph{functional} testing of industrial embedded software,
safety standards call for \emph{structural} testing of the
production code before release deployment.
\rronly{In Sec.~\ref{sec:etester}, we give a brief overview about such
standards and the state of affairs of their implementation in
practice.}

\subsection{Structure of Generated Code}
Many modelling languages follow the \emph{synchronous programming
paradigm}~\cite{Hal93b}, which is well-suited for modelling
time-triggered systems, in which tasks (subsystems of the model)
execute at given rates.
Code generation for such languages produces a typical code structure, which
corresponds essentially to a non-preemptive operating system task scheduler.
Most code generators provide the scheduler for time-triggered execution or
code to interface with popular real-time operating systems.  In either case,
the functionality corresponds to the following pseudo code:

{
\scriptsize
\begin{lstlisting}[numbers=left, xleftmargin=2em]
void main() {
  state s; inputs i; outputs o;
  initialize(s);
  while(true) { //main loop
    i = read_inputs();
    (o,s) = compute_step(i,s);
    write_outputs(o);
    wait(); //wait for timer interrupt
  }
}
\end{lstlisting}
}

The distinguishing characteristic of such a reactive program is
its unbounded main loop, which we will analyse incrementally.  All
other loops contained within that loop, e.g.~to iterate over arrays or
interpolate values using look-up tables, have a statically bounded
number of iterations and can be fully unwound.


\subsection{Analysis with BMC and $k$-induction}\label{sec:etester}

%

\rronly{
\begin{figure}[t]
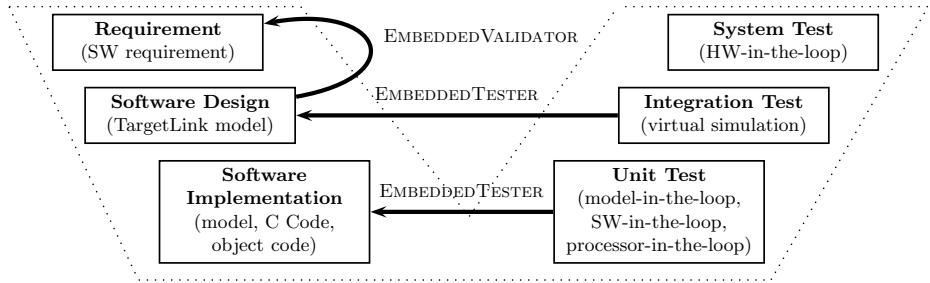

\centering
\scalebox{0.8}{
\small
\begin{tabular}{rrrrrcllllll}
  & \hspace*{3em} & \hspace*{3em} & \hspace*{8em} &\hspace*{4em}& \hspace*{8em} &
  \hspace*{3em} & \hspace*{3em}& \\
  \pnode{ul} &&& \pnode{uil} &&&& \pnode{uir} &&& \pnode{ur}\\
  & \multicolumn{3}{c}{\rnode{req}{\psframebox{\parbox{10em}{\centering \textbf{Requirement}
        \\{\small (SW requirement)}}}}} &&&&
  \multicolumn{3}{c}{\hspace*{2.2em} \psframebox{\parbox{10em}{\centering \textbf{System Test} \\ {\small(HW-in-the-loop)}}}}\hspace*{2.2em} \\[4ex]
  &&
  \multicolumn{2}{c}{\rnode{swd}{\psframebox{\parbox{10em}{\centering
          \textbf{Software Design} \\ {\small(TargetLink model)}}}}} &&&&
  \multicolumn{2}{c}{\rnode{it}{\psframebox{\parbox{10em}{\centering
          \textbf{Integration Test} \\ {\small (virtual simulation)}}}}} \\[4ex]
  &&& \multicolumn{2}{c}{\rnode{swi}{\psframebox{\parbox{10em}{\centering \textbf{Software
            Implementation} \\ {\small (model, C Code, \\ object code)}}}}} & \pnode{c}&
  \multicolumn{2}{c}{\rnode{ut}{\psframebox{\parbox{10em}{\centering
          \textbf{Unit Test} \\ {\small (model-in-the-loop, SW-in-the-loop, processor-in-the-loop)}}}}}\\
  && \pnode{ll} &&&&&& \pnode{lr} 
\end{tabular}
\ncline[arrows=->,linewidth=2pt]{ut}{swi}\nbput{\btcetf}
\ncline[arrows=->,linewidth=2pt]{it}{swd}\nbput{\btcetf}
\nccurve[angleA=10,angleB=10,arrows=->,linewidth=2pt,ncurv=3]{swd}{req}\nbput{\btcevf}
\psset{linestyle=dotted}
\ncline{ul}{uil}
\ncline{ul}{ll}
\ncline{ur}{uir}
\ncline{ll}{lr}
\ncline{lr}{ur}
\ncline{uil}{c}
\ncline{c}{uir}
}
\caption{\label{fig:btcesworkflow}
Embedded software development tool chain in the V model}
\end{figure}
}

\rronly{
Recent safety standards, e.g.~ISO-26262~\cite{ISO26262}, cover model-based
development and testing techniques for early simulation, testing and
verification, and recommend back-to-back
testing\rronly{~\cite{Liggesmeyer2009}} for showing simulation equivalence
between a high-level model and corresponding production code.
In the automotive industry, model-based development including automatic code
generation is well-established.  In particular, \mlsl\rronly{ by The
Mathworks$^\circledR$} for functional modelling and
\dstl\footnote{\url{http://www.dspace.com/en/pub/home/products/sw/pcgs/targetli.cfm}}
for automatic code generation from these models are prominent
representatives.
\textsc{Simulink DesignVerifier}\footnote{\url{http://www.mathworks.com/products/sldesignverifier}},
\btc \btcev and
\btcet\footnote{\url{http://www.btc-es.de/index.php?lang=2&idcatside=14}},
and \textsc{Reactis}\footnote{\url{http://www.reactive-systems.com}}
complement the software development tool chain \rronly{as illustrated in
Fig.~\ref{fig:btcesworkflow} }for formal verification of safety
requirements against design models and re\-qui\-re\-ment-based testing
and back-to-back testing including automatic test vector generation
for structural coverage criteria.
}

\rronly{
\begin{figure}[t]
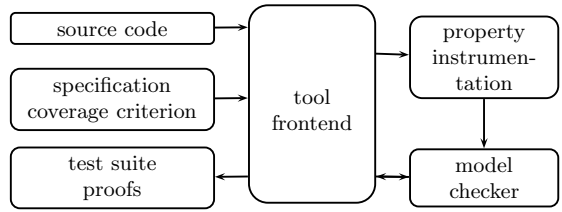

\centering
\scalebox{0.9}{
\psset{framearc=0.3,arrows=->}
\begin{tabular}{r@{\hspace{1.5em}}r@{}c@{}l@{\hspace{1.5em}}l}
& & \multirow{3}{*}{\psframebox{\parbox{5em}{\centering ~\\[7ex] tool \\ frontend \\[7ex]}}}\\[-2ex]
\rnode{sc}{\psframebox{\parbox{8.5em}{\centering source code}}} & 
\Rnode{fe1}{} & & \multirow{3}{*}{\Rnode{fe2}{}} &  
\multirow{3}{*}{\rnode{pi}{\psframebox{\parbox{6em}{\centering property \\ instrumen-\\ tation}}}}\\[2ex]
\multirow{3}{*}{\rnode{sp}{\psframebox{\parbox{8.5em}{\centering specification \\ coverage criterion}}}} & & & & 
 \\
& \Rnode{fe3}{} & &  & \\
& \\
\multirow{3}{*}{\rnode{ts}{\psframebox{\parbox{8.5em}{\centering test suite \\ proofs}}}} & & & & 
\multirow{3}{*}{\rnode{mc}{\psframebox{\parbox{6em}{\centering model \\ checker}}}} \\
& \Rnode{fe5}{} & & \Rnode{fe4}{} \\[2ex]
\end{tabular}
\ncline{sc}{fe1}\ncline{fe2}{pi}
\ncline{sp}{fe3}\ncline{fe4}{mc}\ncline{mc}{fe4}
\ncline{fe5}{ts}
\ncline{pi}{mc}
}
\caption{\label{fig:btcesarch}
Typical architecture of a model checker for embedded software}
\end{figure}
}



\paragraph{Property Instrumentation}
%
%
Formal verification requires formalisations of high-level
requirements, often using observer B\"uchi
automata\rronly{~\cite{Buechi62Decision}} with a
dedicated `error state' generated from temporal logic
descriptions.  Test vector generation is done for code-coverage
criteria such as branches, statements, conditions and
MC/DC~\cite{HVCR01} of the production C code.  For
\fuelsys, for example, MC/DC instrumentation yields $251$ test goals.
The properties to be verified or tested have in common that they can be
reduced to a reachability problem.  In formal verification of safety
properties, we prove that the error state is unreachable, whereas the aim of
test vector generation is to obtain a trace that demonstrates reachability
of the goal state.

To validate whether the air-fuel ratio in the \fuelsys controller is 
regulated after a few seconds to be within some margin of the target ratio, one
has to instrument the reactive program, as sketched above, with an
observer implementing the asserted property.  For instance, consider
the requirement ``If some sensor fails for the first time then within $10$
seconds the air-fuel ratio will keep in between the range of $80\,\%$ to
$120\,\%$ of the target ratio forever.'' The code fragment for
an observer for this requirement may look as follows:

{
\scriptsize
\begin{lstlisting}[numbers=left, xleftmargin=2em]
// detection of first sensor failure
if (sensor_fail == 1 && observe_ratio == 0) {
    // initialize observer variables
    observe_ratio = 1;
    counter = 0;
    violated = 0;
}
\end{lstlisting}
\vspace*{-2ex}
%
%
\begin{lstlisting}[numbers=left, xleftmargin=2em, firstnumber=8]
// if in observation mode...
if (observe_ratio == 1) {
    if (counter >= 10 &&
        (air_fuel_ratio < 0.8*target_ratio ||
         air_fuel_ratio > 1.2*target_ratio)) {
        violated = 1;
    }
    counter++;
}
\end{lstlisting}
\vspace*{-2ex}
\begin{lstlisting}[numbers=left, xleftmargin=2em, firstnumber=17]
// safety property
assert(violated == 0);
\end{lstlisting}
}

In order to verify that the above property actually holds, one has to show
that the assertion in the observer code is always satisfied.  We use BMC for
refutation of the assertion, and $k$-induction for proving it.

\paragraph{Bounded Model Checking}
%
%
BMC \rronly{\cite{BCCZ99,CBRZ01}}\pponly{\cite{BCCZ99}} can be used to check the
existence of a path $\pi=\langle s_0,s_1,\ldots,s_k\rangle$ of
length $k$ between two sets of states described by
propositional formulae $\phi$ to $\psi$.
This check is performed by deciding satisfiability of the following formula
using a SAT or SMT solver:
\begin{equation}\label{equ:bmc}
\phi(s_0)\wedge\bigwedge_{0\leq j<k} \trans(s_j,i_j,s_{j+1}) \wedge \psi(s_k)
\end{equation}
If the solver returns the answer ``satisfiable'', it also provides a
satisfying assignment to the variables $(s_0,i_0,s_1,i_1,\ldots,s_{k-1},i_{k-1},s_k)$.
The satisfying assignment represents one possible path $\pi=\langle
s_0,s_1,\ldots,s_k\rangle$  from $\phi$
to $\psi$  and identifies the corresponding input sequence $\langle
i_0,\ldots,i_{k-1}\rangle$.

Hence, BMC is useful for refuting safety properties (where~$\phi$ gives the
set of initial states and $\psi$ defines the error states) and generating
test vectors (where $\psi$ defines the test goal to be covered).

\paragraph{Unbounded Model Checking by k-Induction}
BMC can prove reachability, whereas unreachability can be shown using
induction.
The predicate $\neg\psi$ is an (inductive) invariant, i.e., it holds
in all reachable states, if each of the following two formulae, base
case (BC) and induction step (SC), are unsatisfiable:
\begin{equation}\label{equ:indinv}
\begin{array}{l@{\qquad}l}
\text{(BC)} & \phi(s) \wedge \psi(s) \\
\text{(SC)} &\neg \psi(s)\wedge T(s,s') \wedge \psi(s')
\end{array}
\end{equation}
Both formulae can be decided with the help of a SAT or SMT solver. 

The property of interest is often not inductive, however, and the
check above fails.  An option is to strengthen the property, e.g.,
using auxiliary invariants obtained using an abstract interpreter.
Furthermore, the criterion above can be generalised to
$k$-induction~\cite{SSS00,ES03b,HT08,DHKR11}:
\begin{equation}\label{equ:kindinv}
\begin{array}{@{\hspace*{-1em} }l@{\quad}l}
\text{(BC)} & \phi(s_0) \wedge \bigwedge_{0\leq j<k}
 \neg\psi(s_j) \wedge \trans(s_j,i_j,s_{j+1}) \wedge \psi(s_k) \\ 
\text{(SC)} & \bigwedge_{0\leq j\leq k}
 \neg\psi(s_j) \wedge \trans(s_j,i_j,s_{j+1}) \wedge
\psi(s_{k+1})
\end{array}
\end{equation}
The base case checks whether $\neg\psi$ holds in the first $k$ steps, 
whereas the induction step checks if we can conclude from the
invariant holding over any $k$ consecutive steps that it holds for the
$(k+1)^{st}$ step.
If the base step fails, i.e.~above formula is satisfiable, we have
refuted the property. If it holds and the induction step fails, we do
not know whether $\neg \psi$ is invariant. Only if both formulae hold
we have proved that $\neg \psi$ is invariant.

Both base step and induction step are essentially instances of
BMC: starting from the initial state $\phi$ for the
base case, and starting from \emph{any} state for the induction step. 
Thus, similar to BMC, $k$-induction can be applied 
by using a sequence of increasing values for $k$.

\paragraph{Challenges}
Embedded C code has to meet many conflicting requirements like
real-time constraints, low memory footprint and low energy
consumption.
Code generators offer options to perform certain optimisations towards
these goals, often to the detriment of \emph{code size} (and also
readability for humans).
The observer instrumentation to encode properties and identify the
test goals corresponding to code-coverage criteria such as MC/DC
produces a non-negligible overhead in the size of the code but
introduces little semantic complexity.
The size of the formula built further increases whenever internal
loops need to be unwound, for example to perform a linear search in
lookup tables in the \fuelsys example.
File sizes of 10\,MB and more are common, which poses difficulties to
many tools already when parsing the source code and encoding the
program into a SAT formula, mostly due to inefficient data structures.
Incremental BMC helps reducing formula sizes and peak memory
consumption (see Sec.~\ref{sec:bench}) by incremental formula
generation and solving.

The next generation of automotive electronic control units will be
equipped with \emph{floating point} (FP) units~\cite{Tricore} to
perform complex arithmetic used to implement control laws. \fuelsys is
an example of this trend.
The use of FP arithmetic comes with many hard-to-avoid pitfalls.
Many \rronly{verification }tools 
interpret FP numbers as reals, an approach which 
gives rise to false positives and is unable to detect certain bugs.
%
Hence, even though bit-precise reasoning about FP arithmetic is
expensive, it is indispensable in order to spot intricate bugs.
State-of-the-art methods are based on bit-blasting and bitvector
refinement \cite{BKW09} (see Sec.~\ref{sec:otherincr}), but new
abstraction-based solvers are emerging
(e.g. \cite{HGBK12}\rronly{\cite{HGBK12,BDG+13}}).

In practice, many loop unwindings may be needed to detect errors and
reach certain tests goals (more than 100 for some of our industrial
benchmarks, see Sec.~\ref{sec:bench}).
\emph{Non}-incremental bounded model checking repeats work such as
file parsing, loop unwinding, SAT formula encoding and discards
information learnt in the SAT solver every time it is called and so
gives away an enormous amount of performance.  This effect is
exacerbating the cost of large unwinding limits that may be needed.

\rronly{
A great challenge is to exploit all the benefits of
incrementality in BMC and to significantly enhance performance of
its integration with an industrial-strength embedded verification and
test-vector generation tool.
}

\section{Incremental BMC}\label{sec:incr}

In this section, we explain the technical background of incremental SAT solving
and how it is employed in our implementation of incremental BMC.

\subsection{Incremental SAT solving}
The first ideas for incremental SAT solving date back to the 1990s
\rronly{\cite{Hoo93,SS97,KWSS00}}\pponly{\cite{Hoo93,SS97}}.  The
question is how to solve a sequence of similar SAT problems while
reusing effort spent on solving previous instances, i.e., reusing the
internal state and learnt information of the solver.

Obviously, incremental SAT solving is easy when the modification to
the CNF representation of the problem makes it grow monotonically.
This means that if we want to solve a sequence of (increasingly
constrained) SAT problems with CNF formulae $\Phi(k)$ for $k\geq 0$
then $\Phi(k)$ must be \emph{growing monotonically} in $k$, i.e.
$\Phi(k+1) = \Phi(k) \wedge \varphi(k)$ for CNF formulae $\varphi(k)$.

Removal of clauses from $\Phi(k)$ is trickier, as some of the clauses
learnt during the solving process are no longer implied by the new
instance, and need to be removed as well.  Earlier work has identified
conditions for the reuse of learnt clauses~\cite{Str01,WKS01}, but
this requires expensive book-keeping, which partially saps the benefit
of incrementality.

The most popular approach to incremental solving is to solve SAT
problems \emph{under assumptions}~\cite{ES03b}: 
assumptions are modelled as the first decision literals made 
by the SAT solver. If a learnt clause is derived from an assumption, then it 
will contain that assumption literal. This light book-keeping enables the 
SAT solver to maintain its performance when using assumptions.
SAT solving under assumptions allows us to emulate the removal 
of clauses as explained in Sec.~\ref{sec:incrbmc}.
SMT solvers offer an interface for pushing and popping
clauses in a stack-like manner. Pushing adds clauses,
popping removes them from the formula. 
This makes the modification of the formula intuitive to the user, 
but the efficiency depends on the underlying implementation of 
the push and pop operations.
For example, in \cite{GW14} it was observed that some SMT solvers (like Z3) are not
optimised for incremental usage and hence perform worse incrementally
than non-incrementally.  Since \textsc{Cbmc} itself implements
powerful bitvector decision procedures, we use the SAT solver \minisat~\cite{ES03a} 
as a backend solver.

We will see that incremental BMC requires 
a \emph{non-monotonic} series of formulae.
For SAT solvers, solving under assumptions is the prevalent method, 
hence we will focus on this technique in the sequel.

\subsection{Incremental BMC}\label{sec:incrbmc}
Following the construction in \cite{ES03b} for finite state machines,
incremental BMC can be formulated as a sequence of SAT problems
$\Phi(k)$ that we need to solve:
\begin{equation}\label{equ:incrbmc}
\begin{array}{rcl}
\Phi(0) &:=& \phi(s_0) \wedge (\Psi(0) \vee \alpha_0) \\
                 && \text{with assumption }\neg \alpha_0\\
\Phi(k+1) &:=& \Phi(k) \wedge \trans(s_k,i_k,s_{k+1})  \wedge \alpha_k
 \wedge \\
   && (\Psi(k+1) \vee \alpha_{k+1}) \\
                 && \text{with assumption } \neg\alpha_{k+1}\\
\end{array}
\end{equation}
where $\Psi(k)$ is the disjunction $\bigvee_{0\leq j\leq k} \psi(s_j)$
of error states $\psi$ to be proved unreachable up to iteration $k$.
This \rronly{disjunction }means that the verification fails if \emph{at least
  one} of the error states is reachable.  Since the set of $\psi_j$s
grows in each iteration, our problem is not monotonic: one has to
\emph{remove} $\Psi(k)$ when adding $\Psi(k+1)$ because $\Psi(k)$
subsumes $\Psi(k+1)$, and thus \rronly{simply }conjoining the $\Psi(k)$s would not
yield the desired formula.

Here, solving under assumptions comes to rescue.
In iteration $k$, the $\alpha_k$ is assumed to be false, whereas
it is assumed true for iterations $k'>k$. This has the effect that 
in iteration $k'$ the formula $(\Psi(k) \vee \alpha_k)$ 
becomes trivially satisfied. Hence, it does not contribute to the
(un)satisfiability of $\Phi(k')$, which emulates 
its deletion.%
\footnote{For a large number of iterations $k$, such
  trivially satisfied subformulas might accumulate as ``garbage'' in the
  formula and slow down its resolution. Restarting the solver 
  at appropriate moments is the common solution to this issue.}

\paragraph{Symbolic execution}
For software, however, (\ref{equ:incrbmc}) results in large
formulae and would be highly inefficient for the purpose of BMC.
In practice, software model checkers use \emph{symbolic execution} in
order to exploit, for example, constant propagation and pruning
branches when conditionals are infeasible, while generating the SAT
formula and thus reducing its size.
This means that the formulae describing $\trans$ and $\Psi$ in
(\ref{equ:incrbmc}) are actually dependent on $k$.
Fortunately, this does not affect the correctness of above formula
construction and we can replace $\trans$ by $\trans_k$ in
(\ref{equ:incrbmc}) and $\psi$ by $\psi_k$ in the definition of
$\Psi(k)$.

\paragraph{Slicing}
Another feature used by state-of-the-art software model checkers is
slicing:
The purpose of slicing is, again, reducing the size of the 
SAT formula by removing (or better: not generating) those parts of the
formula that have no influence on its satisfiability.
There are many techniques how to implement slicing with the desired
trade-off between runtime efficiency and its formula pruning
effectiveness~\cite{Tip94}.

Slicing is performed relative to $\Psi(k)$.  We said that the number
of disjuncts $\psi_j$ in $\Psi$ is growing monotonically with~$k$.
Hence, we will show that, assuming that our slicing operator is
monotonic, we obtain a monotonic formula construction:
The transition relation for each time frame $\trans_k$ is a
conjunction $\bigwedge_{\tau\in M} \tau$ of subrelations $\tau$
(e.g., formulae corresponding to program instructions). The slicing
operator $\slice$ selects a subset of $M$.
The operator $\slice$ is monotonic iff $M \subseteq M' \Longrightarrow \slice(M)
\subseteq\slice( M')$.
We can then view the conjunction of transition relations for $k$ time
frames $\widehat{\trans}(k) = \bigwedge_{0\leq j\leq k} \trans_j$ as
$\bigwedge_{\tau\in M_k} \tau$.  Then a slice 
$\widehat{\trans}^{sliced}(k)$ of $\widehat{\trans}(k)$
is $\bigwedge_{\tau\in M'_k} \tau$ where $M'_k\subseteq M_k$.
An incremental slice is then defined as the difference between  
$\widehat{\trans}^{sliced}(k+1)$ and $\widehat{\trans}^{sliced}(k)$:
\rronly{\begin{equation}
T^{sliced}_{k+1} = \bigwedge_{\tau\in M'_{k+1}\setminus M'_k} \tau.
\end{equation}}
\pponly{$T^{sliced}_{k+1} = \bigwedge_{\tau\in M'_{k+1}\setminus M'_k} \tau$.}
Monotonicity of formula construction follows from $M'_{k+1} \subseteq
M_{k+1}$ and the assumed monotonicity $M'_k \subseteq M'_{k+1}$ of the
slicing operator.  We can thus replace $\trans$ by $\trans^{sliced}_k$
in (\ref{equ:incrbmc}). Mind that $\trans^{sliced}_k$ contains also
subrelations $\tau$ for time steps $k'<k$.

\subsection{Incremental Refinements}\label{sec:otherincr}

Incremental SAT solving is also used for incremental refinements of
the transition relation $\trans$ for bitvectors and arrays,
for example.

\paragraph{Bitvector Refinement}
The purpose of bitvector refinement
\rronly{\cite{BKO+07,Bie08,HH08,BKO+09,BB09c,EMA10}}\pponly{\cite{BKO+07,Bie08,EMA10}} is to reduce the size of formulae
encoding bitvector operations.  This is especially important for
arithmetic operations that generate huge SAT formulae,
e.g. multiplication, division and remainder operations, both for
integer and floating-point variables~\cite{BKW09}.
Bitvector refinement is based on successive under- and
overapproximations.  For instance, underapproximations can be obtained
by fixing a certain number of bits, whereas overapproximation make a
certain number of bits unconstrained.
If an underapproximation is satisfiable (SAT) or an overapproximation is 
unsatisfiable (UNSAT) we
know that the non-approximated formula is SAT or UNSAT
respectively. Otherwise, the number of fixed respectively
unconstrained bits is reduced until the non-approximated formula
itself is checked.

\paragraph{Arrays}
\pponly{To handle programs with arrays, 
Ackermann expansion is necessary to ensure the functional
consistency property of arrays: $\forall i,j: i=j\Longrightarrow A[i]
= A[j]$.  However, adding a quadratic number of constraints (in the size of the
array $A$) is extremely costly. Experience has shown that only a small
number of these constraints is actually used \cite{PS06}. 
A refinement procedure can be used to lazily add
these constraints only when actually necessary.
For details, we refer to SMT solvers, like \textsc{Boolector}, which
implement such a procedure to decide the SMT-LIB array theory
\cite{BB09a,BB09b}.}
\rronly{
To handle programs with arrays, 
Ackermann expansion is necessary to ensure the functional
consistency property of arrays: $\forall i,j: i=j\Longrightarrow A[i]
= A[j]$.  However, adding a quadratic number of constraints (in the size of the
array $A$) is extremely costly. Experience has shown that only a small
number of these constraints is actually used \cite{PS06}. 

Hence, more efficient just trying to solve the SAT formula without
these constraints, which is an over-approximation. Hence, if we get
an UNSAT result (a), we know that the solution with the Ackermann
constraints would be UNSAT too. In case of a SAT result (b), we check
the consistency of the obtained model: if it turns out not to violate
consistency, then we know that we have found a real bug. Otherwise
(c), we add the violated Ackermann constraint to the formula. The
formula construction is trivially monotonic and we can use
incremental SAT solving. We repeat the procedure until we hit case
(a) or (b), which is guaranteed to happen.

Some SMT solvers, like \textsc{Boolector} implement a similar
procedure to decide the SMT-LIB array theory \cite{BB09a,BB09b}.
}

\paragraph{Formula construction}
Applying above refinements inside an incremental Bounded Model Checker
requires using several incremental formula encodings for (in general,
non-monotonic) refinements simultaneously.
These refinements are global over all unwindings, so that in iteration
$k$ we have to further refine transition relations $\trans_{k'}$ from
earlier iterations $k'<k$.
We can formalise the incremental formula construction as follows:
For iteration $k\geq 0$ of incremental BMC and
the~$\ell^{\text{th}}$ refinement:
$$
\begin{array}{r@{\,}c@{\,}l}
\Phi(0,0) &:=& \phi(s_0) \wedge (\Psi_0(s_0) \vee \alpha_0) \\
                 && \text{with assumption }\neg \alpha_0\\
\Phi(k+1,\ell) &:=& \Phi(k,\ell) \wedge (\Psi_{k+1}(s_{k+1}) \vee \alpha_{k+1}) \wedge \alpha_k \wedge\\
              && \big(\trans'_{k+1,\ell}(s_k,i_k,s_{k+1}) \vee \beta_\ell\big) \\
              && \text{with assumptions } \neg\alpha_{k+1}\text{ and
              }\neg\beta_\ell\\
\end{array}
$$
$$
\begin{array}{r@{\,}c@{\,}l}
\Phi(k,\ell+1) &:=& \Phi(k,\ell) \wedge \\
                && \big(\trans'_{k-1,\ell+1}(s_{k-1},i_{k-1},s_k) \vee \beta_{\ell+1}\big) \wedge \beta_\ell \\
              && \text{with assumptions } \neg\alpha_k\text{ and }\neg\beta_{\ell+1}\\
              && \text{for }k\geq 1
\end{array}
$$
$\ell$ is incremented in each iteration of the refinement loop until convergence, whereas $k$ is incremented when considering the next time frame.

\section{Experimental Evaluation}\label{sec:exp}

We present the results of our experimental evaluation of incremental BMC and 
incremental $k$-induction on industrial programs from mainly automotive 
origin. 
The experiments for this study were performed on a 3.5 GHz Intel Xeon machine with 8 
cores and 32 GB of physical memory running Windows 7 with a time limit of 3,600 
seconds.

Our study of incremental BMC is targeted at embedded software since it
takes advantage of its specific properties (one big unbounded loop,
whereas other loops are bounded). However, incremental BMC can also be
applied to programs where loops and control structures are more irregular;
to this end, we report on a preliminary study of incremental BMC on programs with 
multiple loops. These latter experiments were performed on a 3 GHz Intel Xeon with
8 cores and 50 GB of physical memory running Fedora 20 with a time
limit of 3,600 seconds.

\subsection{Implementation}\label{sec:impl}

\rronly{\paragraph{\cbmc}}
We have implemented our extension%
\footnote{Source code available from
\submission{\url{http://drive.google.com/file/d/}\\\url{0B22MA57MHHBKMldKeXFHQTBBZ3M}}
\final{\url{http://www.cprover.org/svn/cbmc/branches/peter-incremental-unwinding}}} 
for incremental BMC in the Bounded Model Checker for ANSI-C programs
\cbmc~\cite{CKL04} using the SAT solver \textsc{Mini}\-\textsc{SAT2}~\cite{ES03a} as a backend solver.

\cbmc is called in incremental mode using the command line 
{\tt cbmc file.c --no-unwinding-assertions --incremental}.\footnote{
  The option \texttt{-{}-no-unwinding-assertions} is required
to prevent \cbmc from inserting an assertion that makes the verification fail
when a loop has not been fully unwound, which is useless in our application
with unbounded loops.
}
The following options can be added to enable specific features of
\cbmc:
\begin{compactitem}
\item \texttt{-{}-no-sat-preprocessor}:
turns off SAT formula preprocessing, i.e. the \minisat simplifier is not used.
\item \texttt{-{}-slice-formula}:
slices the SAT formula.
\item \texttt{-{}-refine}:
 enables bitvector refinement.
\item \texttt{-{}-unwind-max $k$}:
 limits the loop unwindings of the loop
  to be checked incrementally to $k$ unwindings. Without this option,
  \cbmc will not terminate for unsatisfiable instances,
  i.e. bug-free programs with unbounded loops.
\end{compactitem}

More information regarding the usage of incremental \cbmc can be found 
on the CPROVER wiki page.\footnote{\url{http://www.cprover.org/wiki/doku.php?id=how_to_use_incremental_unwinding}}

\rronly{\paragraph{Integration with an industrial-strength embedded
  verification tool}}
In the integration of \cbmc with \btc \btcet and \btcev, a master
routine selects the next verification/test goal to be analysed
starting from instrumented C code.  After some preprocessing like
source-level slicing and internal-loop unwinding the resulting
reachability task is given to \cbmc.  If \cbmc is able to solve the
problem within the user-defined time limit, the result,
i.e.~information of bounded or unbounded unreachability, or a test
vector or counterexample in case of reachability, is reported back to
the master process.  Otherwise, i.e.~in case of a timeout, \cbmc is
killed but information about the solved unwindings of the reactive
main loop is given back, which frequently is a useful result for the
user.

To prove unreachability of verification/test goals (properties), split-case
$k$-induction is performed (see Section~\ref{sec:etester}). For this
purpose \btc \btcet generates two source files, one containing the
base case, which is a normal BMC problem with the property given as
assertion (cf. Equ. (\ref{equ:kindinv}) (BC)); the file for the step
case havocs variables modified in the loop and the invariant property
is assumed at the beginning of the loop and asserted at the end of the
loop (cf. Equ. (\ref{equ:kindinv}) (SC)). To check the step case, we
require a reversed termination behaviour of \textsc{Cbmc} (option
\texttt{-{}-stop-when-unsat}), i.e. it continues unwinding as long as
the problem is SAT and stops as soon as it is UNSAT.

\rronly{\paragraph{Implementation of Incremental BMC for General Sequential 
Programs}}

Incremental \cbmc can also be used for programs with multiple
loops. For these programs, \cbmc incrementally unwinds loops one at
each time.  For a given loop, \cbmc will unwind the loop until it is
fully unwound or until a maximum depth $k$ is reached (given by option
{\tt --unwind-max $k$}).  After a loop has been unwound, \cbmc
continues to the next loop. This procedure is repeated until all loops
have been unwound or a bug has been found.  Recursive function calls 
are treated similarly.

\begin{table*}[t]
\centering
{\scriptsize
\begin{tabular}{|ll|r|rrr|rrr|rrr|r|r|}
\hline
 & & & \multicolumn{3}{c|}{operators} & \multicolumn{3}{c|}{input variables} & \multicolumn{3}{c|}{state variables} & observer &  \\
& & LOC & cond & mul & div/rem & bool & int & float & bool & int & float & bool & unwindings\\
\hline
\multirow{2}{*}{SAT} & max & 
31222 &	17103 & 669 & 75 & 688 & 477 & 189 & 3876 & 750 & 107 & 22 & 106 \\
& average & 7572 & 4306 & 188 & 9 & 103 & 79 & 19 & 583 & 136 & 15 & 9 & 22 \\
\hline
\multirow{2}{*}{UNSAT} & max & 
23014 & 49530 & 567 & 37467 & 212 & 282 & 188 & 708 & 663 & 32 & 22 & 10 \\
& average &
4854 & 6014 & 160 & 1257 & 30 & 51 & 9 & 163 & 73 & 3 & 7 & 10 \\
\hline
\end{tabular}}~\\[2ex]
\caption{\label{tab:benchmarksum} Benchmark characteristics from industrial programs}
\end{table*}

\subsection{Incremental BMC for Embedded Software}\label{sec:bench}

We report results on industrial programs for the integration of 
\cbmc with \btc \btcet and \btcev.
For these experiments, we used 60 benchmarks that originated mainly 
from automotive applications.\footnote{Unfortunately, 
the source code of the benchmarks 
cannot be made public due to non-disclosure agreements.} 
Half of the benchmarks are bug-free (UNSAT instances), half contain a
bug (SAT instances).
This benchmark suite is an indicator for performance of model checking tools
in practice as it covers a representative spectrum of embedded software.  
A~summary of the benchmark characteristics is listed in
Table~\ref{tab:benchmarksum}; for a full list we refer to
Table~\ref{tab:benchmarks} in the Appendix.

Besides the number of lines of code, we give the
number of conditional operators, multiplications and divisions or 
remainder operations, which are a good indicator for the
difficulty of the benchmark, because they generate large
formulae --- recall that for each ``\texttt{/}'' occurring in the program,  
\cbmc~has to generate a divider circuit.
The surprisingly high number of conditional operators in most of the
benchmarks is due to the preprocessing of conditional assignments
by \btc \btcet and hints at the amount of branching in these benchmarks.
Moreover, we list the number of input and state variables, and the
variables introduced by the observer instrumentation.

These benchmarks have the property of having only one unbounded loop. 
For these benchmarks, \cbmc is called in incremental mode by 
using the option {\tt --incremental-check c:main.0} where {\tt c::main.0} is 
the loop identifier of the unbounded loop to be unwound and checked incrementally. 
The loop identifiers can be obtained using the option {\tt --show-loops}.


\begin{figure}[t]
\centering
   \begin{subfigure}{0.43\linewidth} 
    \subcaptionbox{Effect of slicing, SAT formula preprocessing and 
    bitvector refinement}[5.2cm]{\hspace*{-1em}
     \includegraphics{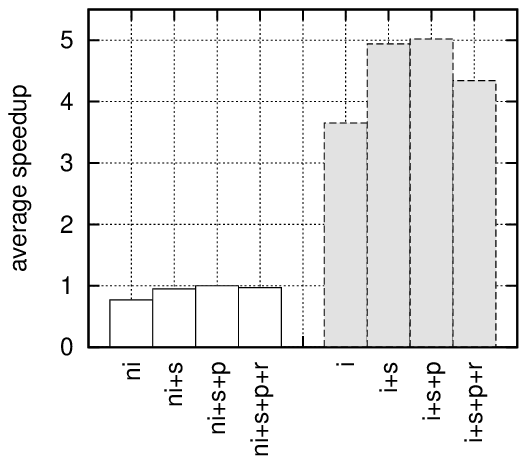}\vspace*{3ex}
     }
   \end{subfigure}
   \begin{subfigure}{0.5\linewidth} 
   \subcaptionbox{Comparison between ni+s+p and i+s+p ($+$ SAT instances; 
     $\Box$ UNSAT instances)}[6.2cm]{\hspace*{-1.5em}
     \includegraphics[scale=0.75]{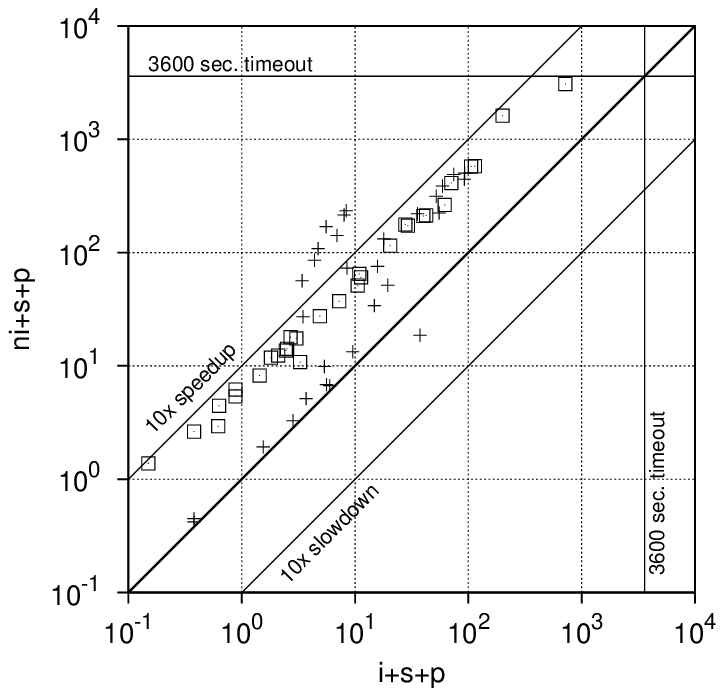}
     }
     
   \end{subfigure}
\caption{Incremental vs. non-incremental BMC} \label{fig:all}
\end{figure}

\paragraph{Runtimes}\label{sec:res}
We compared the incremental (i) with the non-incremental (ni) approach and
evaluated the impact of slicing (s), SAT preprocessing (p) 
and bitvector refinement (r).%
\footnote{Array refinement is not used because the
benchmarks do not contain arrays.}
The incremental and non-incremental approaches were compared by 
activating none of the three techniques, with slicing only (+s), with
slicing and preprocessing (+s+p), and with all three options activated
(+s+p+r).

The maximum number of loop unwindings was fixed to 10 for the UNSAT instances 
in order to balance a significant exploration depth with reasonable analysis 
runtimes. For SAT instances, a maximum number of loop unwindings was not fixed 
since the incremental and non-incremental approaches are bound to terminate 
when the unwinding depth reaches the depth of the bug.
The number of unwindings are listed in the last column in
Table~\ref{tab:benchmarksum} (resp. Table~\ref{tab:benchmarks} in the Appendix).

Fig.~\ref{fig:all} shows the comparison between the incremental and 
non-incremental approaches and the impact of each tool option on 
their performance.

Fig.~\ref{fig:all}a shows the average geometric mean~\cite{JW86} speedup
of instances that were solved by all approaches.
%
%
We consider as baseline the (ni+s+p) approach since it was the best
non-incremental approach. Each bar shows the average geometric mean
speedup of each approach when compared to (ni+s+p). For example, (ni)
has a speedup of 0.77, i.e. (ni) is on average 0.77$\times$ slower
than (ni+s+p).
%
On the other hand, all incremental versions are much faster than the
non-incremental versions. For example, (i) is on average over
3.5$\times$ faster than (ni+s+p) and (i+s+p) is on average over
5$\times$ faster than (ni+s+p).
We observe the following effects of the tool options:
(i) slicing shows significant benefits overall (also on peak memory
consumption); (ii) not using formula preprocessing is a bad idea in
general; and (iii) bitvector refinement shows benefits for UNSAT
instances, but produces overhead for SAT instances which deteriorates
the overall performance of the tool (see Fig.~\ref{fig:all-sat-unsat}
in the Appendix for more details).
Even though the tool options have some positive effects, they are
rather minor in comparison to the performance gains from using an
incremental approach.

Since the best incremental and non-incremental approaches were
obtained with the configuration (+s+p), we will use this configuration
for both approaches on the results described in the remainder of the
paper.

Fig.~\ref{fig:all}b shows a scatter plot with runtimes of the best
non-incremental (ni+s+p) and incremental (i+s+p) approaches. Each
point in the plot corresponds to an instance, where the x-axis
corresponds to the runtime required by the incremental approach and
the y-axis corresponds to the runtime required by the non-incremental
approach. If an instance is above the diagonal, then it means that the
incremental approach is faster than the non-incremental approach,
otherwise it means that the non-incremental approach is faster.
SAT instances are plotted as crosses, whereas UNSAT instances are
plotted as squares.
Incremental BMC significantly outperforms non-incremental BMC.  For
SAT instances, the advantage of incremental BMC is negligible for the
easy instances, whereas speedups are around a factor of 10 for the
medium and hard instances.
For UNSAT instances, speedups are also significant and 
most instances have a speedup of more than a factor of 5.

\paragraph{Solving vs. overall runtime}
Since the non-incremental approach has to re-parse source files and
preprocess them at each iteration, one might argue that removing this
overhead is the main reason for the speedup observed. However, the overhead 
for parsing files, symbolic execution and slicing when compared to generating 
and solving SAT formula is similar for the incremental and non-incremental 
approach. 
The incremental approach spends 27\% of its time solving the SAT formula 
(582 out of 2,151 seconds), whereas the non-incremental 
approach spends 28\% of its time (3,317 out of 11,811 seconds).
Unsurprisingly, solving the instance for the largest $k$ in the non-incremental 
approach takes a considerable amount of time (around 24\%), 
when compared to the total time for solving the SAT formulae for iterations 1\ldots k 
(784 out of 3,317 seconds). 

An explanation for these speedups might be the size of the queries
issued in both approaches. The average number of clauses per solver
call is halved from 1,367k clauses for the non-incremental approach to 
709k clauses for the incremental approach. Similarly, the average number of 
variables is less than a third in the incremental approach when compared to the 
non-incremental approach, being of 217k and 746k respectively.







\rronly{\paragraph{Peak memory consumption}}
Smaller query sizes also have an effect on peak memory consumption
which is reduced by 30\% for UNSAT benchmarks; for SAT benchmarks,
however, we observed a 10\% increase.

\subsection{Incremental BMC for Code coverage on {\sc FuelSys}}
As reported in the previous section, enabling \cbmc to work
incrementally led to tremendous performance gains\rronly{ on the benchmark
suite consisting of selected single input files}.  In order to assess
whether these improvements have
practical impact in the \emph{integration} of \cbmc with an
industrial-strength test-vector generation tool, we compared the
performance of \submission{the industrial verification
  tool}\final{\btc \btcet} with the incremental feature of \cbmc being
disabled and enabled.  The time limit per subtask was $10$~minutes and
the unwinding depth for all internal loops was $50$.
For unwinding depth~$10$ of the main loop, the incremental feature
improves the overall runtime from $152.3$ to $70.4$~minutes,
i.e.~more than 2$\times$ faster, and for unwinding depth~$50$ from
$377.4$ to $108.5$~minutes, i.e.~more than 3$\times$ faster.
\rronly{In the latter case, the rate of solved subproblems for MC/DC
  (i.e.\ not run into timeout) could be increased from $98.4\,\%$ to
  $99.2\,\%$. 
}

\subsection{Incremental $k$-Induction for Embedded Software}

To compare the performance of incremental and non-incremental approaches for 
$k$-induction, we considered the subset of UNSAT benchmarks for which 
$k$-induction required more than 1 iteration. 
Note that when $k$-induction requires only 
1 iteration, the performance of incremental and non-incremental approaches is 
similar. 
This subset of benchmarks corresponds to 10 UNSAT benchmarks (see 
Table~\ref{tab:benchmarks} in the Appendix for more details).

Fig.~\ref{fig:kind} shows a scatter plot with the runtimes of incremental 
and non-incre\-men\-tal $k$-induction using the tool options (+s+p).
Instances that correspond to the base case are plotted as crosses, whereas 
instances that correspond to the step case are plotted as squares.
The runtimes for both incremental and non-incremental checking are
relatively small. These are due to the small number of iterations
required by $k$-induction to prove the unreachability of the
properties present on these benchmarks (between 2 and 4 iterations
with an average of 2.4 per instance).

Incremental checking is always faster than non-incremental checking. 
When considering the average geometric mean speedup, 
incremental checking is around 2$\times$ faster than non-incremental checking, 
on both base and step cases.

\begin{figure}[!t]
\pponly{\vspace*{-4ex}}
\centering
\begin{minipage}{.5\textwidth}
\hspace*{-4em}
  \includegraphics[scale=0.77]{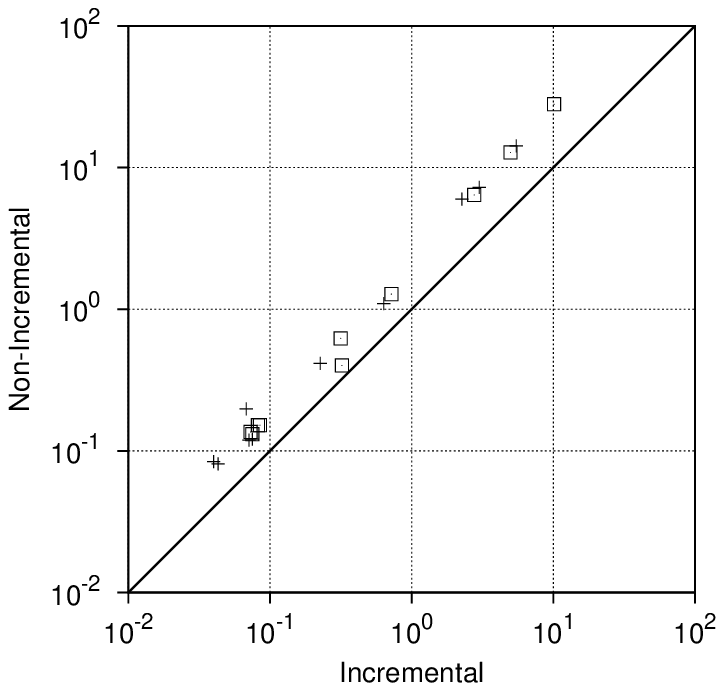}
  \captionof{figure}{Incremental $k$-induction\\
  ($+$ base case; $\Box$ step case)}
  \label{fig:kind}
\end{minipage}%
\begin{minipage}{.5\textwidth}
\hspace*{-3em}
  \includegraphics[scale=0.78]{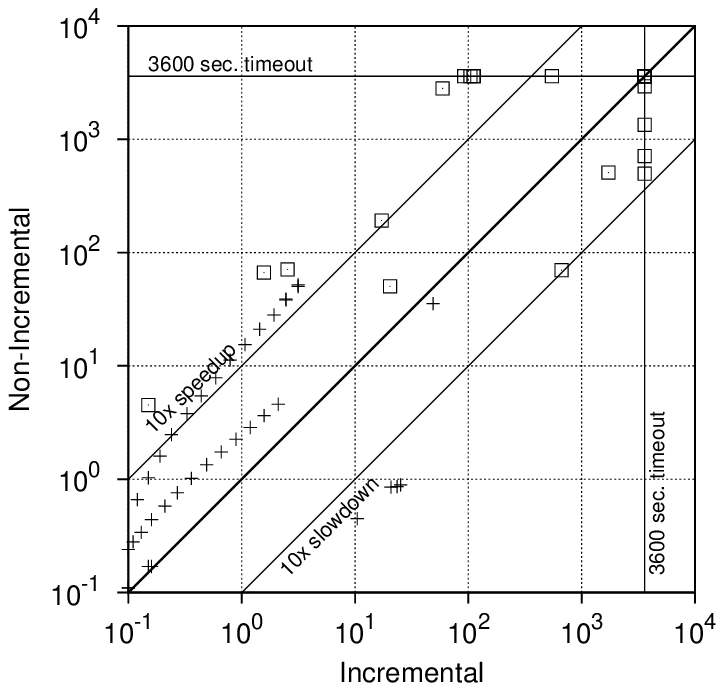}
  \captionof{figure}{Incremental BMC for SystemC\\
  ($+$ SAT instances; $\Box$ UNSAT instances)}
  \label{fig:systemc}
\end{minipage}
\end{figure}

\subsection{Incremental BMC for Programs with Multiple Loops}\label{sec:gen}

Incremental BMC is not restricted to programs with a single unbounded loop 
and may also be applied to programs with multiple unbounded loops. 
To evaluate the performance of incremental BMC on this kind of
programs, we compared the performance of incremental and
non-incremental approaches on the 62 benchmarks from the SystemC category
of the Software Verification Competition benchmark set,%
\footnote{\rronly{Available at
  }\url{https://svn.sosy-lab.org/software/sv-benchmarks/trunk/c/systemc/}}
because these benchmarks, which were derived from SystemC
models~\cite{MR10}, contain many loops.
25 benchmarks are bug-free (UNSAT instances) and 37 contain a bug (SAT
instances).
These benchmarks have between 2 and 19 loops with an average of 10.3
loops per instance. For SAT instances, the depth of the bug ranges
from 1 to 5 with an average depth of 2.5. For more details regarding
these benchmarks see Table~\ref{tab:benchmarks-systemc} in the
Appendix.

We have fixed the maximum number of loop unwindings to 10 for both,
SAT and UNSAT instances. Note that this unwind depth is larger than
the depth of the bugs for the SAT instances.
Formula slicing is not yet fully supported in incremental \cbmc for
programs with multiple loops. Therefore, the incremental approach was
run with the tool options (+p), whereas the non-incremental approach
was run with the tool options (+s+p).

Fig.~\ref{fig:systemc} shows a scatter plot with the run times of the 
incremental and non-incremental approaches. 
For the majority of the instances, the incremental approach
outperforms the non-incremental approach and for many SAT and UNSAT
instances the speedup is larger than a factor of 10.  However, there
are a few instances for which the non-incremental approach performs
better.
The non-incremental approach unwinds all loops until a fixed unwind
depth, whereas the incremental approach fully unwinds one loop before
continuing to the next loop. For some instances, fully unwinding each
loop may result in the generation of larger formulae, particularly for
SAT instances.
Not using slicing for the incremental approach may also result in larger 
formulae. The increase in formula size may explain the observed slowdown 
for some instances.

\rronly{Overall, when considering instances solved by both approaches, 
the incremental approach is faster than the non-incremental approach 
and the average geometric speedup is larger than a factor of 3.}

\pponly{Overall, when considering instances solved by both approaches, 
the incremental approach is on average 3$\times$ faster than the 
non-incremental approach.}

\section{Related Work}\label{sec:rw}

Most related is recent work on a prototype tool \textsc{nbis}
\cite{GW14} implementing incremental BMC using SMT solvers. They show 
the advantages of incremental software BMC.
However, they do not consider industrial embedded software and 
have evaluated their tool only on small benchmarks
that are very easy for both, incremental and non-incremental, approaches
(runtimes $<$1s).%
\footnote{Unfortunately, a working version of the tool was not available at time of submission.}

Bit-precise formal verification techniques are indispensable 
for embedded system models and implementations, 
that have low-level, i.e. C language, semantics
like discrete-time \simulink models.
The importance of this topic has recently attracted attention as shown
by publications on verification using SMT Solving \cite{HRB13,MMBC11}, test
case generation \cite{PRS+12}, symbolic analysis for improving
simulation coverage \cite{AKRS08}, and directed random testing
\cite{SYR08}. Yet, all these works have not exploited incremental
BMC.

The test vector generation tool \textsc{FShell} \cite{HSTV09} uses
incremental SAT solving to check the reachability of a set of test
goals. However, it assumes a fixed unwinding of the loops.
There is no reason why incremental BMC should not
boost its performance when increasing loop unwindings need to be
considered.
Test vector generation tools like \textsc{Klee} \cite{CDE08}
use incremental SAT solving to extend the paths to be explored.
However, they consider only single paths at a time, whereas
BMC explores all paths simultaneously.

Incremental SAT solving has important applications in other
verification techniques like the IC3 algorithm \cite{Bra12a,EMB11} and
incremental BMC is standard for hardware
verification~\cite{JS05,Wie11}.  We show that the speedups of
incremental SAT solving reported in~\cite{ES03b} regarding
$k$-induction on small HW circuits carry over to industrial embedded
software.

\section{Conclusions}\label{sec:concl}

We claim that incremental BMC is an indispensable technique 
for embedded software verification that should be considered
state-of-the-art in such tools.
To underpin this claim, we report on the successful integration of our
incremental extension of \textsc{Cbmc} into an industrial embedded
software verification tool. Our experiments demonstrate
one-order-of-magnitude speedups from incremental approaches on
industrial embedded software benchmarks for BMC and $k$-induction.
These performance gains result in faster property verification and
higher test coverage, and thus, a productivity increase in embedded
software verification.

Moreover, we reported on significant speedups for programs with
multiple loops that show the applicability of incremental BMC beyond
embedded software.
We expect that incremental BMC can be further improved for programs
with multiple loops by simultaneously unwinding all loops
incrementally instead of fully unwinding one loop at each
time. Incremental $k$-induction for programs with multiple
loops will also benefit from such an improvement.

\small
\bibliographystyle{splncs03}
\bibliography{biblio}

\newpage
\appendix
\section{Industrial Benchmark Characteristics}
\vspace*{-5ex}
\begin{table}[hb!]
\centering
\scalebox{0.66}{
\begin{tabular}{|l|r|rrr|rrr|rrr|r|r|}
\hline
 & & \multicolumn{3}{c|}{operators} & \multicolumn{3}{c|}{input variables} & \multicolumn{3}{c|}{state variables} & observer &  \\
name & LOC & cond & mul & div/rem & bool & int & float & bool & int & float & bool & unwindings\\
\hline
automotive\_sat\_01 & 3762 & 2032 & 82 & 1 & 14 & 282 & 0 & 229 & 50 & 0 & 3 & 12 \\
automotive\_sat\_02 & 1854 & 189 & 79 & 1 & 78 & 4 & 0 & 165 & 7 & 0 & 3 & 15\\
automotive\_sat\_03 & 15277 & 17103 & 669 & 75 & 230 & 244 & 0 & 868 & 275 & 0 & 1 & 9 \\
automotive\_sat\_04 & 13853 & 16908 & 601 & 59 & 208 & 219 & 0 & 741 & 266 & 0 & 1 & 12\\
automotive\_sat\_05 & 469 & 193 & 90 & 11 & 1 & 0 & 0 & 17 & 3 & 0 & 3 & 21\\
automotive\_sat\_06 & 10702 & 5117 & 646 & 1 & 7 & 54 & 19 & 28 & 60 & 22 & 16  & 5 \\
automotive\_sat\_07 & 10970 & 5068 & 646 & 1 & 7 & 54 & 19 & 27 & 62 & 22 & 15  & 4\\
automotive\_sat\_08 & 3656 & 2657 & 79 & 1 & 14 & 61 & 26 & 20 & 68 & 30 & 16 & 2 \\
automotive\_sat\_09 & 253 & 34 & 79 & 1 & 0 & 3 & 0 & 23 & 4 & 0 & 3 & 103\\
automotive\_sat\_10 & 604 & 117 & 79 & 1 & 23 & 7 & 0 & 81 & 10 & 0 & 3 & 40\\
automotive\_sat\_11 & 592 & 115 & 79 & 1 & 23 & 7 & 0 & 79 & 10 & 0 & 3 & 48\\
automotive\_sat\_12 & 1978 & 2201 & 79 & 1 & 0 & 0 & 0 & 4 & 172 & 0 & 3 & 53\\
automotive\_sat\_13 & 1980 & 2198 & 79 & 1 & 0 & 0 & 0 & 4 & 172 & 0 & 3 & 55\\
automotive\_sat\_14 & 1222 & 216 & 79 & 1 & 0 & 26 & 0 & 94 & 67 & 0 & 3 & 56\\
automotive\_sat\_15 & 5020 & 3172 & 79 & 1 & 18 & 4 & 0 & 115 & 22 & 0 & 3 & 17\\
automotive\_sat\_16 & 2578 & 4572 & 89 & 4 & 1 & 20 & 105 & 3 & 22 & 107 & 17 & 2\\
automotive\_sat\_17 & 2580 & 4592 & 89 & 4 & 1 & 20 & 105 & 2 & 22 & 107 & 18 & 1\\
automotive\_sat\_18 & 2740 & 4718 & 89 & 4 & 1 & 20 & 105 & 2 & 24 & 107 & 16 & 2\\
automotive\_sat\_19 & 27456 & 3579 & 177 & 7 & 546 & 95 & 0 & 3426 & 438 & 0 & 1 & 12\\
automotive\_sat\_20 & 27456 & 3579 & 177 & 7 & 546 & 95 & 0 & 3426 & 438 & 0 & 1 & 16\\
automotive\_sat\_21 & 31222 & 3705 & 178 & 7 & 688 & 477 & 0 & 3876 & 750 & 0 & 1 & 12\\
automotive\_sat\_22 & 30834 & 3620 & 177 & 7 & 652 & 476 & 0 & 3837 & 744 & 0 & 1 & 14\\
automotive\_sat\_23 & 1270 & 508 & 102 & 5 & 6 & 66 & 0 & 79 & 124 & 9 & 16 & 1\\
automotive\_sat\_24 & 1272 & 501 & 102 & 5 & 6 & 66 & 0 & 78 & 124 & 9 & 17 & 3\\
automotive\_sat\_25 & 1282 & 506 & 102 & 5 & 6 & 67 & 0 & 79 & 128 & 9 & 15 & 1\\
automotive\_sat\_26 & 321 & 28 & 79 & 1 & 6 & 2 & 0 & 36 & 2 & 0 & 3 & 106\\
avionics\_sat & 2214 & 1413 & 79 & 2 & 30 & 16 & 0 & 189 & 52 & 0 & 1 & 20\\
fuelsys\_sat\_01 & 9402 & 16603 & 311 & 6 & 0 & 0 & 4 & 31 & 5 & 8 & 22 & 1\\
fuelsys\_sat\_02 & 9404 & 16757 & 311 & 6 & 0 & 0 & 4 & 31 & 5 & 8 & 22 & 1\\
fuelsys\_sat\_03 & 5746 & 8521 & 224 & 3 & 0 & 0 & 4 & 30 & 5 & 7 & 19 & 1\\
automotive\_unsat\_01* & 3761 & 2032 & 82 & 1 & 14 & 282 & 0 & 229 & 50 & 0 & 3 & 10 \\
automotive\_unsat\_02 & 3762 & 2032 & 82 & 1 & 14 & 282 & 0 & 229 & 50 & 0 & 3 & 10 \\
automotive\_unsat\_03 & 1579 & 889 & 79 & 1 & 0 & 38 & 0 & 75 & 4 & 0 & 3 & 10 \\
automotive\_unsat\_04 & 1853 & 189 & 79 & 1 & 78 & 4 & 0 & 165 & 7 & 0 & 3 & 10 \\
automotive\_unsat\_05 & 503 & 321 & 106 & 19 & 1 & 0 & 0 & 21 & 3 & 0 & 3 & 10 \\
automotive\_unsat\_06 & 13259 & 16672 & 545 & 59 & 188 & 207 & 0 & 708 & 232 & 0 & 1 & 10 \\
automotive\_unsat\_07 & 464 & 193 & 90 & 11 & 1 & 0 & 0 & 17 & 3 & 0 & 3 & 10 \\
automotive\_unsat\_08 & 23014 & 49530 & 536 & 37467 & 92 & 220 & 0 & 697 & 304 & 0 & 1 & 10 \\
automotive\_unsat\_09 & 4768 & 3334 & 79 & 1 & 0 & 26 & 0 & 215 & 663 & 0 & 3 & 10 \\
automotive\_unsat\_10 & 1035 & 160 & 79 & 1 & 30 & 4 & 0 & 115 & 29 & 0 & 1 & 10 \\
automotive\_unsat\_11 & 12142 & 5859 & 567 & 0 & 7 & 54 & 19 & 27 & 60 & 22 & 17 & 10 \\
automotive\_unsat\_12 & 12518 & 6242 & 567 & 0 & 7 & 54 & 19 & 27 & 62 & 22 & 15 & 10 \\
automotive\_unsat\_13* & 4726 & 3091 & 42 & 0 & 14 & 61 & 26 & 30 & 71 & 32 & 16 & 10 \\
automotive\_unsat\_14* & 591 & 115 & 79 & 1 & 23 & 7 & 0 & 79 & 10 & 0 & 3 & 10 \\
automotive\_unsat\_15* & 1977 & 2198 & 79 & 1 & 0 & 0 & 0 & 4 & 172 & 0 & 3 & 10 \\
automotive\_unsat\_16  & 2339 & 559 & 82 & 9 & 22 & 56 & 0 & 170 & 79 & 0 & 3 & 10 \\
automotive\_unsat\_17* & 1399 & 258 & 79 & 1 & 0 & 29 & 0 & 106 & 73 & 0 & 3 & 10 \\
automotive\_unsat\_18* & 5021 & 3172 & 79 & 1 & 18 & 4 & 0 & 115 & 22 & 0 & 3 & 10 \\
automotive\_unsat\_19* & 7979 & 12127 & 119 & 15 & 0 & 0 & 0 & 5 & 16 & 0 & 3 & 10 \\
automotive\_unsat\_20* & 6217 & 686 & 88 & 2 & 212 & 87 & 0 & 697 & 60 & 0 & 1 & 10 \\
automotive\_unsat\_21* & 5230 & 1043 & 81 & 2 & 99 & 24 & 0 & 511 & 112 & 0 & 1 & 10 \\
automotive\_unsat\_22 & 190 & 97 & 90 & 11 & 0 & 0 & 0 & 4 & 31 & 0 & 1 & 10 \\
automotive\_unsat\_23 & 659 & 93 & 79 & 1 & 9 & 1 & 0 & 75 & 10 & 0 & 3 & 10 \\
automotive\_unsat\_24 & 3554 & 787 & 81 & 52 & 16 & 79 & 0 & 226 & 45 & 0 & 3 & 10 \\
automotive\_unsat\_25 & 1575 & 184 & 79 & 1 & 38 & 0 & 0 & 199 & 15 & 0 & 3 & 10 \\
avionics\_unsat & 2329 & 1413 & 79 & 2 & 30 & 16 & 0 & 188 & 52 & 0 & 1 & 10 \\
fuelsys\_unsat\_01* & 5146 & 17271 & 214 & 5 & 0 & 0 & 3 & 11 & 0 & 5 & 21 & 10 \\
fuelsys\_unsat\_02 & 7806 & 19764 & 215 & 6 & 0 & 0 & 4 & 31 & 5 & 8 & 22 & 10 \\
fuelsys\_unsat\_03 & 7804 & 19764 & 215 & 5 & 0 & 0 & 4 & 31 & 5 & 8 & 22 & 10 \\
fuelsys\_unsat\_04 & 3340 & 11671 & 205 & 3 & 0 & 0 & 3 & 15 & 0 & 3 & 18 & 10 \\
\hline
\end{tabular}}
\caption{\label{tab:benchmarks} \small Embedded software benchmark characteristics 
(name of the benchmark and application domain, lines of code, number of
  operators (cond(\texttt{a?b:c}), mul(\texttt{*}), div/rem(\texttt{/},\texttt{\%})), 
  number of boolean/integer/floating point
  input and state variables, number of boolean variables introduced by
  the observer instrumentation, number of loop unwindings considered; 
  k-induction was performed on the instances marked with *) }
\end{table}

\newpage
\section{Additional Results for BMC Comparison}


\begin{figure}[h]
\centering
   \begin{subfigure}{0.48\linewidth} \centering
    \subcaptionbox{SAT benchmarks}[5.2cm]{
     \includegraphics{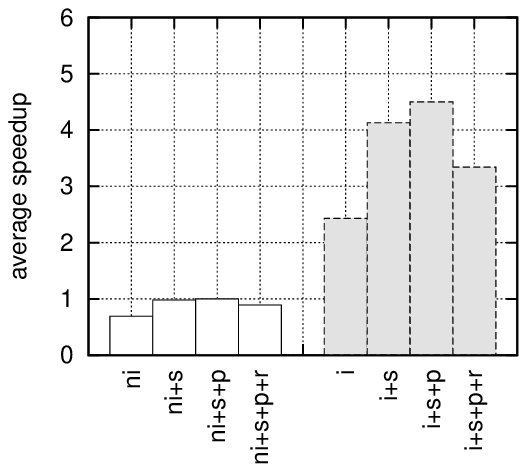}
     }
   \end{subfigure}
   \begin{subfigure}{0.48\linewidth} \centering
   \subcaptionbox{UNSAT benchmarks}[6.2cm]{
     \includegraphics{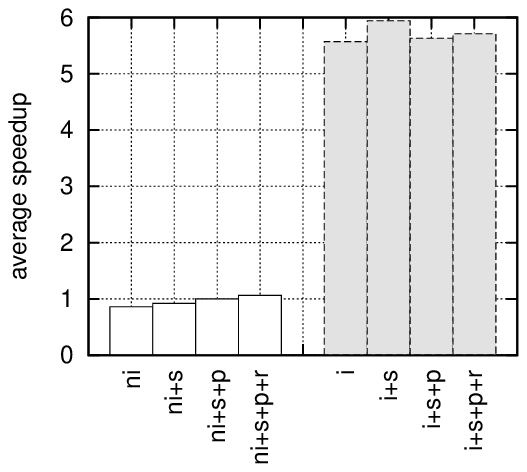}
     }
     
   \end{subfigure}
\caption{Effect of slicing, SAT formula preprocessing and 
    bitvector refinement} \label{fig:all-sat-unsat}
\end{figure}

\newpage
\section{SystemC Benchmark Characteristics}

\begin{table}[hb!]
\centering
\scalebox{0.64}{
\begin{tabular}{|l|r|rr|r|}
\hline
& & \multicolumn{2}{c|}{loops} & \\
name & LOC & bounded & unbounded & unwindings \\
\hline
bist\_cell\_sat.cil.c & 240 & 0 & 2 & 10 \\
kundu\_sat.cil.c & 290 & 3 & 2 & 10  \\
mem\_slave\_tlm.1\_sat.cil.c & 724 & 0 & 13  & 10  \\
mem\_slave\_tlm.2\_sat.cil.c & 729 & 0 & 13  & 10  \\
mem\_slave\_tlm.3\_sat.cil.c & 734 & 0 & 13  & 10  \\
mem\_slave\_tlm.4\_sat.cil.c & 739 & 0 & 13  & 10  \\
mem\_slave\_tlm.5\_sat.cil.c & 744 & 0 & 13  & 10  \\
pc\_sfifo\_1\_sat.cil.c  & 172 & 0 & 4 & 10  \\
pc\_sfifo\_2\_sat.cil.c  & 214 & 0 & 4 & 10  \\
pc\_sfifo\_3\_sat.cil.c  & 258 & 0 & 4 & 10  \\
pipeline\_sat.cil.c  & 400 & 0 & 3 & 10  \\
token\_ring.01\_sat.cil.c & 210 & 0 & 5 & 10  \\
token\_ring.02\_sat.cil.c & 270 & 0 & 6 & 10  \\
token\_ring.03\_sat.cil.c & 330 & 0 & 7 & 10  \\
token\_ring.04\_sat.cil.c & 390 & 0 & 8 & 10  \\
token\_ring.05\_sat.cil.c & 450 & 0 & 9 & 10  \\
token\_ring.06\_sat.cil.c & 510 & 0 & 10  & 10  \\
token\_ring.07\_sat.cil.c & 570 & 0 & 11  & 10  \\
token\_ring.08\_sat.cil.c & 630 & 0 & 12  & 10  \\
token\_ring.09\_sat.cil.c & 690 & 0 & 13  & 10  \\
token\_ring.10\_sat.cil.c & 750 & 0 & 14  & 10  \\
token\_ring.11\_sat.cil.c & 810 & 0 & 15  & 10  \\
token\_ring.12\_sat.cil.c & 870 & 0 & 16  & 10  \\
token\_ring.13\_sat.cil.c & 930 & 0 & 17  & 10  \\
toy\_sat.cil.c & 315 & 0 & 6 & 10  \\
kundu1\_unsat.cil.c  & 233 & 2 & 2 & 3 \\
kundu2\_unsat.cil.c  & 285 & 3 & 2 & 2 \\
pc\_sfifo\_1\_unsat.cil.c  & 173 & 0 & 4 & 1 \\
pc\_sfifo\_2\_unsat.cil.c  & 215 & 0 & 4 & 1 \\
pipeline\_unsat.cil.c  & 400 & 0 & 3 & 5 \\
token\_ring.01\_unsat.cil.c & 217 & 0 & 5 & 3 \\
token\_ring.02\_unsat.cil.c & 277 & 0 & 6 & 3 \\
token\_ring.03\_unsat.cil.c & 337 & 0 & 7 & 3 \\
token\_ring.04\_unsat.cil.c & 397 & 0 & 8 & 3 \\
token\_ring.05\_unsat.cil.c & 457 & 0 & 9 & 3 \\
token\_ring.06\_unsat.cil.c & 517 & 0 & 10  & 3 \\
token\_ring.07\_unsat.cil.c & 577 & 0 & 11  & 3 \\
token\_ring.08\_unsat.cil.c & 637 & 0 & 12  & 3 \\
token\_ring.09\_unsat.cil.c & 697 & 0 & 13  & 3 \\
token\_ring.10\_unsat.cil.c & 757 & 0 & 14  & 3 \\
token\_ring.11\_unsat.cil.c & 817 & 0 & 15  & 3 \\
token\_ring.12\_unsat.cil.c & 877 & 0 & 16  & 3 \\
token\_ring.13\_unsat.cil.c & 937 & 0 & 17  & 3 \\
token\_ring.14\_unsat.cil.c & 875 & 0 & 16  & 3 \\
token\_ring.15\_unsat.cil.c & 935 & 0 & 17  & 3 \\
toy1\_unsat.cil.c  & 317 & 0 & 6 & 3 \\
toy2\_unsat.cil.c  & 314 & 0 & 6 & 3 \\
transmitter.01\_unsat.cil.c  & 197 & 0 & 6 & 2 \\
transmitter.02\_unsat.cil.c  & 256 & 0 & 7 & 2 \\
transmitter.03\_unsat.cil.c  & 315 & 0 & 8 & 2 \\
transmitter.04\_unsat.cil.c  & 374 & 0 & 9 & 2 \\
transmitter.05\_unsat.cil.c  & 433 & 0 & 10  & 2 \\
transmitter.06\_unsat.cil.c  & 492 & 0 & 11  & 2 \\
transmitter.07\_unsat.cil.c  & 551 & 0 & 12  & 2 \\
transmitter.08\_unsat.cil.c  & 610 & 0 & 13  & 2 \\
transmitter.09\_unsat.cil.c  & 669 & 0 & 14  & 2 \\
transmitter.10\_unsat.cil.c  & 728 & 0 & 15  & 2 \\
transmitter.11\_unsat.cil.c  & 787 & 0 & 16  & 2 \\
transmitter.12\_unsat.cil.c  & 846 & 0 & 17  & 2 \\
transmitter.13\_unsat.cil.c  & 905 & 0 & 18  & 2 \\
transmitter.15\_unsat.cil.c  & 905 & 0 & 18  & 1 \\
transmitter.16\_unsat.cil.c  & 961 & 0 & 19  & 1 \\
\hline
\end{tabular}}
\caption{\label{tab:benchmarks-systemc} \small SystemC benchmark characteristics 
(name of the benchmark, lines of code, number of bounded loops, number of 
  unbounded loops, and number of loop unwindings considered)}
\end{table}

\end{document}